\documentclass[letterpaper, 10 pt, conference]{ieeeconf}  

\IEEEoverridecommandlockouts                              
\title{\LARGE \bf
Design of an Optoelectronic State Machine with integrated BDD based Optical logic
}

\author{Macauley Coggins
}

\usepackage[T1]{fontenc}
\usepackage{graphicx}

\begin{document}

\maketitle
\thispagestyle{empty}
\pagestyle{empty}

\begin{abstract}

In this paper I demonstrate a novel design for an optoelectronic State Machine which replaces input/output forming logic found in conventional state machines with BDD based optical logic while still using solid state memory in the form of flip-flops in order to store states. This type of logic makes use of waveguides and ring resonators to create binary switches. These switches in turn can be used to create combinational logic which can be used as input/output forming logic for a state machine. Replacing conventional combinational logic with BDD based optical logic allows for a faster range of state machines that can certainly outperform conventional state machines as propagation delays within the logic described are in the order of picoseconds as opposed to nanoseconds in digital logic.    

\end{abstract}
\section{INTRODUCTION}

Throughout the 20th and early 21st century it can be seen that state machines as a whole have been purely electronic as is the same with most classical  computing. Most classical state machines are composed of digital logic in the form of NAND or NOR based logic which has allowed state machines to be easily designed and implemented in to many electronic systems that require sequential logic including elevators, vending machines, and traffic lights, as well as in computer systems. As we are reaching the end of Moore's law many groups in the industry are funding research in to finding potential candidate solutions that will outperform current digital implementations. Optolectronic systems are an example of a non-classical candidate as the use of light as a medium for transferring and manipulating information has been shown to be much faster than current all digital computing[1]. Unlike purely optical systems (which would certainly outperform any optoelectronic system but so far have proven difficult to design and implement for use in computing) optoelectronic systems are also more viable in the short term given that they can make use of digital components such as solid state memory. Optoelectronic state machines are an example of an optoelectronic system that would outperform its digital counterpart as input/output forming logic would be handled optically using waveguides and ring resonators (rather than electronically using transistors) but whose states are stored conventional using solid state memory. 

\section{Conventional State Machines}

A conventional state machine can be defined as a system made up of a memory storage medium which contains the state of the system (normally in the form of flip-flops as they have two accessible states each normally denoted as Q and $\bar{Q}$ ) and input/output forming logic (which is made up on discrete combinational logic) in order to access a finite set of states based upon a number of inputs. Using this system the machine can progress from one state to another with the next state depending on the inputs as well as the present state.[2] A conventional state machine can be categorized in to two models. One is the Mealy machine whose outputs depend on both the current state and inputs. The other is the Moore machine which differs in that output depends on the current state only. 

\section{Overall Architecture of a Optoelectronic State Machine}

\begin{figure}
	\includegraphics[width=\linewidth]{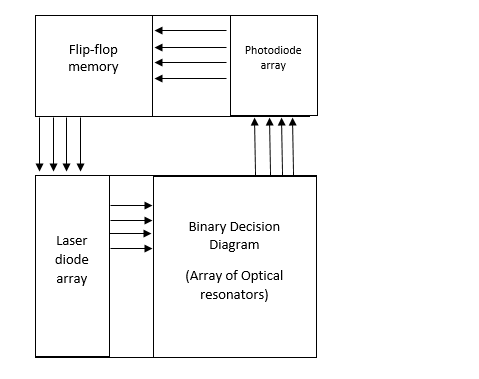}
	\caption{Overall Architecture.}
\end{figure}

\subsection{Summary}

The main architecture of the optoelectronic state machine can be simplified in to 4 main blocks. The first block contains the digital flip-flops which will hold the states of the system. The second block can be defined as the laser diode array interconnection between the first block and second block is done electronically. This block contains the laser diodes which are driven by the output pins of the first block and are used to pump the input ring resonators found in the third block. The third block is the optical input/output forming logic. The block as a whole is purely optical and contains the waveguides and input ring resonators. The resonators act as binary switches and when pumped with light from the laser diodes in the second block the switches will allow light to pass through them in to another waveguide therefore routing the light in the circuit. Output from the optical logic will pass through output waveguides that make up the interconnection between block three and block four. Block four contains the photodiode array. These photodiodes are coupled to the output waveguides from block three and produce a current when light intensity is sufficient. Current from the photodiode array will in turn drive the J/K pins of the flip-flops found in the first block. 

\subsection{State Input and Memory}

The main memory of the optoelectronic state machine described in this paper is similar to memory found in conventional state machines in that flip flop memory is implemented. The main differences however lie in the interface between the optical logic and the memory as any output from the logic will be optical and as such will require conversion from light in to electrical current with an array of photodiodes in order to change the states found in the memory. This electrical current in turn asserts the J or K pin of a JK flip-flop and as such will change the state of the machine. Interconnection between the output of the logic to the photodiode array is made up of a waveguides from the output directly coupled with the photodiodes. Such implementations have shown to offer a high bandwidth of >40Ghz as well as a quantum efficiency of 70\%[3].

\subsection{State Output and BDD based logic using optical resonators}

Input in to the optical BDD logic must first be converted from electrical current to light. This is done by driving laser diodes with the Q or $\bar{Q}$ pins of each flip flop. This will allow for specific sections of logic to be light pumped depending on output conditions of each flip flop. 

The input/output forming logic of the optoelectronic state machine uses a Binary Decision Diagram based system which is implemented using optical ring resonators. Ring resonators are a set of waveguides with at least one loop. They are used in many applications such as filtering due to the fact that ring resonators filter out wavelengths that are not at resonance within the closed loop. In previous experiments ring resonators have been used for optical switching [4]. These have been fabricated using electron beam lithography on a SOI layer which consists of a 0.3$\mu{m}$ Si top layer and a 1$\mu{m}$ buried oxide layer. Ring resonators have been fabricated with a radius of 5$\mu{m}$ and a width of 0.4$\mu{m}$. Waveguides and ring resonators were fabricated 0.2$\mu{m}$ apart [5]. Use of optical resonators is a much easier way of creating logic for use in optical computing given their simplicity and the fact that they can be used as binary switches. Multiple ring resonators and messengers can be used to form combinational logic. Optical resonator based BDD logic works by introducing an optical signal in to an input waveguide coupled to an optical ring resonator with a modulation depth of 10 dB such that when a laser with a wavelength of 532 nm pumps the optical resonator it will switch the resonance of the ring and thus the output of the input waveguide. Light from the messenger will come from a laser that is coupled to the input of the input waveguide using lens fiber. Any waveguides whose output terminals will be unused may require an additional design that will reduce reflection at the output terminal. 

\begin{figure}[h]
  \includegraphics[scale=0.5]{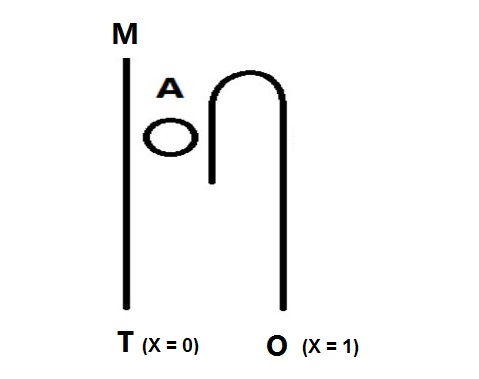}
  \caption{Schematic of an optical resonator implemented binary switch}
\end{figure}

\subsection{Example logic}

\subsubsection{Inverter}

One of the simplest logic circuits we can demonstrate is an inverter which requires only 2 waveguides and one optical resonator in order to function. The output of the gate (denoted as f(x) in fig. 3.) is the output of the input waveguide and will only output light when resonator A is not optically pumped. When resonator A is pumped however light from the input waveguide will flow through resonator A and output out of the other waveguides terminal.

\begin{figure}[h]
	\includegraphics[scale=0.5]{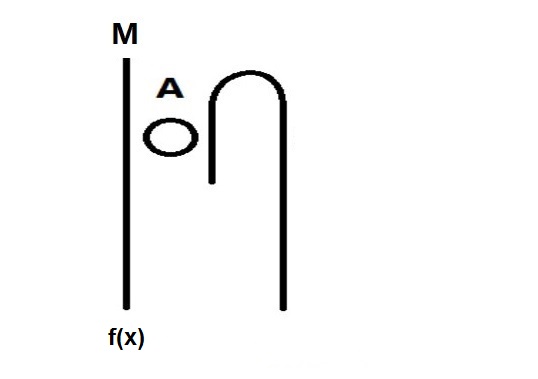}
	\caption{Schematic of an inverter using Optical Resonators}
\end{figure}

\subsubsection{AND gate}
An AND gate which includes three waveguides and two optical resonators denoted in fig. 4. as A and B. Light going through the input waveguide will only output from the f(x) channel when both A and B resonators are optically pumped with a 532nm laser. 

\begin{figure}[h]
  \includegraphics[scale=0.5]{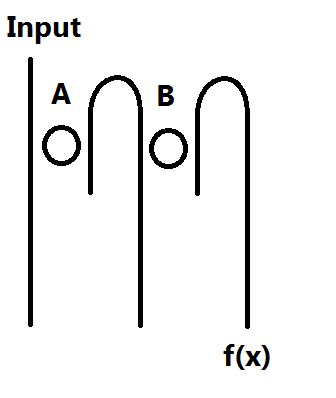}
  \caption{Schematic of a AND gate using Optical Resonators}
\end{figure}

\subsubsection{XOR gate}

An XOR gate will require 3 optical resonators acting as two different outputs. In fig. 5. it can be seen that input A has one optical resonator. Input B has two optical resonators which will both be pumped from the same laser. If either (but not both) inputs are pumped then the light will output through $C_0$. If both inputs are pumped through their respective optical resonators then light will output through the f(x) channel. 

\begin{figure}[h]
  \includegraphics[scale=0.8]{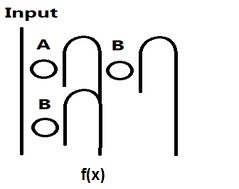}
  \caption{Schematic of an XOR gate using Optical Resonators}
\end{figure}

\section{Example implementations of Optoelectronic State Machines}

\subsection{A Simple 4 bit up Counter}

One example of a implemented optoelectronic state machine is a 4 bit up counter. The design in this paper consists of 4 JK Flip Flops and 2 optical AND gates which use the optical resonator based BDD logic described in Section 3 Subsection C. Figure 6 shows the schematic for the counter described. It can be seen that the Q pin of each Flip-Flop (excluding $O_D$) drives a Laser Diode. These will then pump light in to a wave guide which will then be pumped in to optical resonators that make up the optical logic. For example light from $O_A$ will pump $A_0$ and $O_B$ will pump $B_0$. Some designs (such as this counter) may require output from one optical gate to be pumped in to the input of another. However light from a gates output may not be suitable for pumping an optical resonator as it may not have the correct wavelength needed to reliably pump the resonator. As such light from the output maybe used to saturate a waveguide photodiode which in turn will drive a laser diode which has a more suitable output wavelength of 532nm. Such an implementation can be seen in Figure 6 in the case of $C_0$. Note that laser sources for input waveguides are not shown in any of the diagrams for simplicity. 

\begin{figure}[h]
  \includegraphics[scale=0.7
  ]{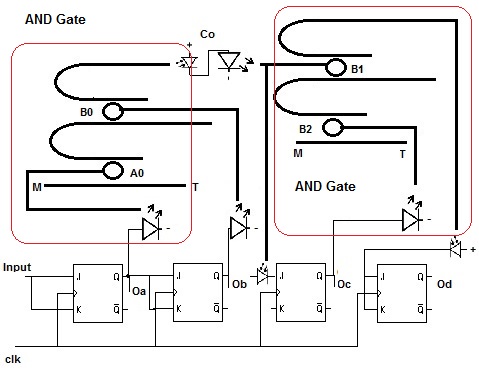}
  \caption{Schematic of a 4 bit Optoelectronic Counter}
\end{figure}

\subsection{11011 Sequence Detector}

A more advanced example is a sequence detector that detects the binary sequence 11011. In Figure 7 it can be seen that such a detector can be designed using only 3 flip flops as well as 4 gates to form the input/output forming logic. The overall design follows the same general architecture as the 4 bit up counter shown in Figure 5 where the Q pin from a Flip-Flop may be used to drive a laser diode and in turn the inputs of the input/output forming logic. The J and K pins may be driven by the optical logic with the use of a photodiode that will be saturated by light from the output of a optical gate and thus drive the J or K pin of a Flip-Flop.   

\begin{figure}[h]
  \includegraphics[scale=0.5
  ]{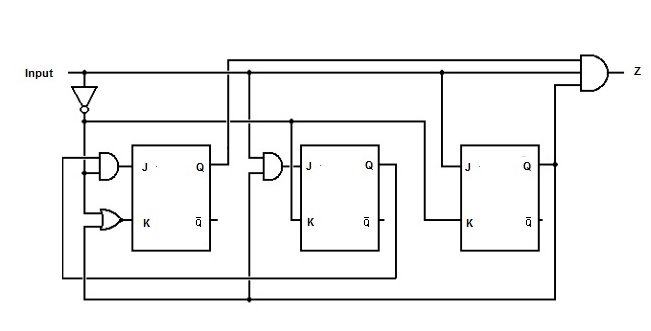}
  \caption{Schematic of a Conventional 11011 sequence detector}
\end{figure}

\begin{figure}[h]
  \includegraphics[scale=0.5
  ]{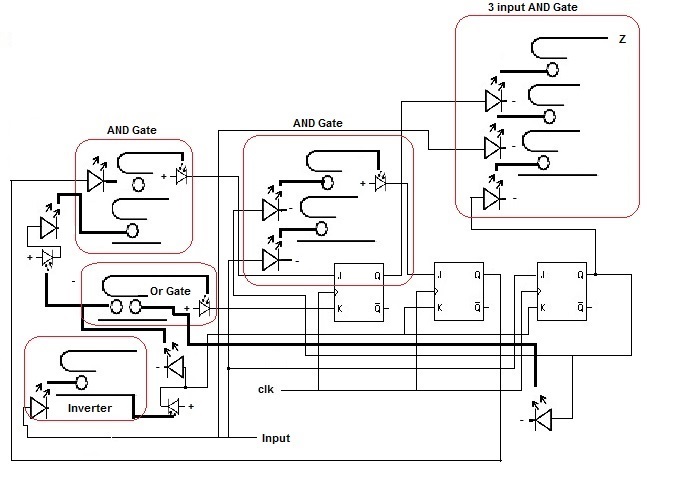}
  \caption{Schematic of an Optoelectronic 11011 sequence detector}
\end{figure}

\section{CONCLUSIONS}

To conclude this paper has shown a novel design for a optoelectronic state machine that would outperform conventional state machines in terms of speed. Two examples of this design have been shown that have real world applications as well as examples of BDD based optical logic used to model logic gates (see Fig. 3-5). These optical logic circuits are what allow for high performance for the designed state machine as a whole as such circuits have a propagation delay that is within picoseconds as opposed to conventional digital logic which may have a propagation delay within nanoseconds. This paper has also shown how light signals between optical logic gates can be used to drive another laser source using a waveguide photodiode coupled with said laser source such that light going in to the input of the next optical circuit will have a wavelength of 532nm that is more suitable for pumping input ring resonators found in the optical circuit. As such this allows for combinational logic circuits to be built from optical logic gates which in turn can be used as input/output forming logic for use in state machines.

\addtolength{\textheight}{-12cm}

\end{document}